\documentclass[prb,showpacs]{revtex4}
\usepackage{amsmath}
\usepackage{amssymb}
\usepackage{graphicx}
\usepackage{graphics}
\usepackage{bm}

\begin{document}

\title{ The role of excitons and trions on electron spin polarization in
quantum wells.}
\author{P. Aceituno}
\email{paceitun@ull.es}
\author{A. Hern{\'{a}}ndez-Cabrera}
\email{ajhernan@ull.es}
\affiliation{Dpto. F{\'{\i}}sica B\'{a}sica. Universidad de La Laguna. La Laguna.
38206-Tenerife. Spain}
\date{\today }

\begin{abstract}
We have studied the time evolution of the electron spin polarization under
continuous photoexcitation in remotely $n$-doped semiconductor quantum
wells. The doped region allows us to get the necessary excess of free
electrons to form trions. We have considered electron resonant
photoexcitation at free, exciton and trion electron energy levels. Also, we
have studied the relative effect of \ photoexcitation energy density and
doping concentration. In order to obtain the two-dimensional density
evolution of the different species, we have performed dynamic calculations
through the matrix density formalism. Our results indicate that
photoexcitation of free electron level leads to a higher spin polarization.
Also, we have found that increasing the photoexcitation energy or
diminishing the doping enhances spin polarization.
\end{abstract}

\maketitle

\section{Introduction}

In semiconductor heterostructures, and in an environment with an excess of
electrons or holes, not only neutral excitons ($X$) can be found but also
negative or positively charged excitons, respectively. This happens because,
under certain circumstances, excitons capture an extra charge to form the so
called trions or charged excitons. Thus, trions represent bound states of
three particles: two electrons and a hole in the case of negatively charged
excitons ($X^{-}$), and two holes and one electron in the case of positively
charged excitons ($X^{+}$). Bound complexes of three particles have a
binding energy which is large enough to make them observable. The most
broadly used experimental technique to investigate trions is the
time-resolved photoluminescence  \cite{1,2,3}. Recently, generation and
recombination processes of excitons and trions have been analyzed using this
technique \cite{4,5}. Photoluminescence has also been used to investigate
two-dimensional electron gas (2DEG) in magnetic fields \cite{6}.

A trion can be formed by\ photoexcitation together with selectively doping
to get the charge excess or, alternatively, only by injection or diffusion
of electrons and holes. There are three ways to form trions in
photoexcitation. Directly, by the resonant excitation of trion electron
level and the trapping of a second electron (or hole) . Indirectly, either
by the excitation of an electron-hole pair, followed by the formation of an
exciton, or by resonant excitation of a neutral exciton. In the latter two
cases the exciton, after its formation, is attached to an existing free
electron (or hole). The dynamics of the three processes are different  \cite{7}.
The other method involves the injection of electrons and holes by
doping regions close to the structure. Injection has the advantage of
avoiding the interaction between the electromagnetic field associated with
the photoexcitation and the excited electrons  \cite{8,9}.

Spintronics has newly aroused great interest in the scientific community due
to its promising future for creating circuits faster and more efficient than
that existing in semiconductor devices. To paraphrase Awschalom et al., the
reason is that the energy needed to generate and transport electron spins is
much less than that necessary to create charge electron currents. In
practice, the main obstacle that faces spintronics is to produce polarized
spin currents without loss of polarization during the process \cite{10}.

One of the main properties of trions is precisely their influence on the
polarization and duration of the electron spin in semiconductor quantum
wells (QWs) and, thus, in spintronics \cite{11,12}. At zero magnetic field
the ground state of the trion is a singlet, that is, the two electrons (or
holes) have opposite spins. The trion triplet state, in which the two
electrons (holes) have the same spin direction, is not bound at zero
magnetic field. Triplet state only becomes bound under finite magnetic
fields. As mentioned above, the trion state consists of two charges
belonging to the most abundant group and the remaining charge belonging to
the minority of the total carrier concentration. Both types of carriers may
have two directions of angular momentum (in a typical quantum well, $\pm \
3/2$ for heavy holes and $\pm \ 1/2$ for electrons). In the singlet state
the total angular momentum coincides with the angular momentum of the
minority carriers, either holes in $X^{-}$ or electrons in $X^{+}$. The
triplet state is observed only in high magnetic fields or when the Zeeman
splitting is greater than the energy separation between singlet and triplet
states \cite{13,14,15,6,16}. Since its simplicity, the singlet state is
particularly suitable for the study of the spin dynamics.

In the last years several works have considered photoexcitation and spin
polarization in different semiconductor structures and under different
resonance conditions. Dynamic studies of spin polarization, including
exciton and trion contributions, can be found in the literature \cite%
{17,18,19,20}. Some authors have reported the resonant photoexcitation at
the exciton or trion electron levels is the suitable way to get the spin
coherence during a reasonably long time to be used in spintronic devices,
without the help of applied magnetic fields or tunneling from magnetic
impurities \cite{21,22}. Others say it is better resonant photoexcitation of
free electron level \cite{23}. However, still is open the question about the
relation between resonance and efficiency. To be precise, what happen if the
energy of photoexcitation resonates with the free electron level, the
exciton level, or the trion level.\ If the excitation energy is in resonance
with one level or another, the situation will be different. In this work we
will compare the three choices just mentioned and show that the resonant
free electron excitation leads to a higher spin polarization, by means of
the intermediate contribution of excitons and trions. We will consider a
long duration pulse to study its effect on the spin polarization.

Suppose we have a\ quantum well doped in the left barrier, as Fig. 1 shows.
Applying a suitable low electric field we can vary the relative position of
the Fermi level with respect to the ground level of the conduction well,
controlling the electron density injected by tunneling into the well. When
we photoexcite free electrons coherently, if the excitation pulse is long
enough, free electrons will form excitons through the Coulomb interaction
with remaining holes. In addition, as the material is slightly doped with
donors, the appearance of negative trions is an immediate consequence. Note
that, in quantum wells, the strong dipole-dipole repulsion between excitons
prevents the formation of biexcitons. Thus, contributions from these neutral
species can be overlooked  \cite{24}. Thus, we deal with three different
electronic states (free, excitonic, and trionic electron) which will occupy
different energy levels.


While free electrons from doping have their spins randomly oriented, it does
not happen the same for photoexcited electrons if illuminating with
polarized light: photoexcited species have a dominant spin orientation
controlled by the polarization of the photon. For example, applying a
circularly polarized light $\sigma ^{+}$, holes are spin up orientated
whilst electrons are spin down \cite{25}. Because of this preferred
orientation one can achieve a net electron spin polarization for practical
purposes. The aim of this paper is the theoretical analysis of the temporal
evolution of the spin polarization for electrons when semiconductor QW is
doped with different concentrations and photoexcited at different laser
frequencies, corresponding to free electron, exciton or trion resonance.
Besides, we will analyze the duration of the spin polarization depending on
the relationship between doping and photoexcitation energy density. We will
consider all the possible processes of generation and annihilation of
excitons, trions and free electrons. In general, we will not consider in
detail the specific physical processes that lead to different relaxation
times (Dyakonov-Perel, Elliott-Yafet, Bir-Aronov-Pikus mechanisms, and so
on). These processes have been carefully analyzed elsewhere \cite{11,12}.
Therefore, we chose empirical relaxation times for the simplest case, where
there are no external fields (except a very low electric field necessary to
inject the free electrons into the QW), inhomogeneities in the interfaces,
interaction with the nuclear spin, etc.

We will show that, although the presence of trions is essential to preserve
a reasonable spin polarization, photoexcitation of electrons at trion level
is not necessary. We use the matrix density formalism, considering the
different generation, recombination and annihilation rates for different
types of binding energy and spin polarization \cite{26}. The effect of the
interband optical pulse is included through the interband generation
function.

We will first consider the Bloch equations for the general dynamics of the
spin polarization for the different species. Then, we will consider the
three cases mentioned before, corresponding to the photoexcitation energy in
resonance with free electron, exciton or trion level.

\section{General dynamics}

We assume an ultrafast \ $\delta (t)$ injection of electrons from the doped
region into the QW to simplify calculations, when the electronic level lays
under the Fermi level. This assumption can be removed if selective doping is
deemed. We consider the excitation laser pulse by the function 
\begin{equation}
w(t)=\frac{1+\cosh (\tau _{p}/2\tau _{f})}{\left[ \cosh ((t-t_{0})/\tau
_{f})+\cosh \left( \tau _{p}/2\tau _{f}\right) \right] },  \label{1}
\end{equation}%
where $\tau _{p}$ is the pulse duration and $10\tau _{f}$ is the front pulse
duration. Thus, for $\tau _{f}=0$ ps we have a square pulse and, for $\tau
_{f}>\tau _{p}$, the pulse in a gaussian-like one \cite{26}. We simulate a
continuous regime excitation by a very long pulse with $\tau _{p}=2000$ ps
and $\tau _{f}=10$ ps. To consider a pulse starting at $t=0$ ps, we have
introduced the shift $t_{0}=\left( \tau _{p}+10\tau _{f}\right) /2$.

We first consider the general quantum kinetic equation for the density
matrix operator $\widehat{\eta }(t)$, and for electrons placed in an
electric field with frequency $\omega $, $\mathbf{E}(t)\exp (-i\omega
t)+c.c.=\mathbf{E}w(t)\exp (-i\omega t)+c.c.$, 
\begin{equation}
\frac{\partial \widehat{\eta }(t)}{\partial t}+\frac{i}{\hbar }\left[ 
\widehat{H},\widehat{\eta }(t)\right] =\frac{1}{i\hbar }\left[ \left( 
\widehat{\delta H_{t}}\exp \left( -i\omega t\right) +H.c.\right) ,\widehat{%
\eta }(t)\right] .  \label{2}
\end{equation}%
Here $\widehat{H}$ is the QW one-particle many-band Hamiltonian we have
described elsewhere \cite{27}. When electrons are excited by the transverse
electric field associated to the laser pulse, $E_{\bot }w(t)\exp (-i\omega
t)+c.c.$, the perturbation operator $\widehat{\delta H_{t}}$ , which
describes interband transition, can be written as \cite{28}%
\begin{equation}
\widehat{\delta H}_{t}=(ie/\omega )E_{\bot }\hat{v}_{\bot }w(t),  \label{3}
\end{equation}%
where $\hat{v}_{\bot }$ is the transverse velocity operator. If we project
on the conduction band states we find the kinetic equation for the
one-electron density matrix $\widehat{\rho }(t)$ 
\begin{equation}
\frac{\partial \widehat{\rho }(t)}{\partial t}+\frac{i}{\hbar }\left[ 
\widehat{H},\widehat{\rho }(t)\right] =\widehat{G}(t)+\widehat{J}\left( 
\widehat{\rho }|t\right) ,  \label{4}
\end{equation}%
where $\widehat{J}\left( \widehat{\rho }|t\right) $ is the collision
integral and the generation rate is given by 
\begin{equation}
\widehat{G}(t)=\frac{1}{\hbar ^{2}}\int_{-\infty }^{0}d\tau e^{\lambda \tau
-i\omega \tau }\left[ e^{i\hat{H}\tau /\hbar }\left[ \widehat{\delta H}%
_{t+\tau },\hat{\rho}_{eq}\right] e^{i\hat{H}\tau /\hbar },\widehat{\delta H}%
_{t}^{+}\right] +H.c.  \label{5}
\end{equation}%
with the phenomenological constant $\lambda \rightarrow +0$. This constant
is the finite relaxation rate. Here $\hat{\rho}_{eq}$ is the equilibrium
density matrix when the second-order contributions to the response are taken
into account.

Taking the basis $\widehat{H}\left\vert \phi _{\alpha }\right\rangle
=\varepsilon _{\alpha }\left\vert \phi _{\alpha }\right\rangle $, we can
rewrite Eq. (3) \ as a system of kinetic equations for $f_{\alpha \beta
}(t)=\left\langle \phi _{\alpha }|\widehat{\rho }(t)|\phi _{\beta
}\right\rangle $, where we neglect non-diagonal terms if $\left( \varepsilon
_{\alpha }-\varepsilon _{\beta }\right) /\hbar $ are larger than the
collision relaxation and generation rates. Thus, for the diagonal terms,%
\begin{equation}
\frac{\partial f_{\alpha \alpha }(t)}{\partial t}=G_{\alpha }(t)+J(f_{\alpha
\alpha }|t),  \label{6}
\end{equation}%
where $G_{\alpha }(t)$ is the photogeneration rate for the $\alpha $ state,
and $J(f_{\alpha \alpha }|t)$ is the collision integral rate. Using the
dipole approximation and the basis $\left\vert \phi _{\alpha }\right\rangle
=\left\vert n_{\alpha }\right\rangle $, with $\alpha =v$ (valence), $c$
(conduction), we get%
\begin{eqnarray}
\frac{dn_{c}(t)}{dt} &=&G(t)+\left( \frac{\partial n_{c}}{\partial t}\right)
_{sc},  \notag \\
\frac{dn_{v}(t)}{dt} &=&-G(t)+\left( \frac{\partial n_{v}}{\partial t}%
\right) _{sc},  \label{7}
\end{eqnarray}%
where $n_{v,c}(t)$ are the electron densities in the valence and conduction
bands. For the case in which there is only photoexcitation, terms $\left( 
\frac{\partial n_{c}}{\partial t}\right) _{sc}=-\left( \frac{\partial n_{v}}{%
\partial t}\right) _{sc}$correspond to the collision-induced relaxation of
population in the conduction and valence bands. If we define the detuning
energy as $\Delta =\hbar \omega -(\varepsilon _{c}-\varepsilon _{v})$, the
interband generation rate can be expressed through%
\begin{equation}
G(t)=2\left( \frac{eE_{\bot }v_{cv}}{\hbar \omega }\right) ^{2}w(t)\it{Re}
\left[ \int_{-\infty }^{0}d\tau w(t+\tau )\sin \left( \frac{\Delta \tau }
{\hbar }\right) \right] ,  \label{8}
\end{equation}%
and $v_{cv}$ is the interband velocity. We will use the reduced generation
function $g(t)=G(t)/N_{ph}$ where%
\begin{equation}
N_{ph}=2\pi \rho _{2D}\left[ eE_{\bot }v_{cv}\left\langle \phi
_{c}(z_{e})|\phi _{v}(z_{h})\right\rangle /\omega \right] ^{2}\tau _{p}/\hbar
\label{9}
\end{equation}%
is the characteristic density of photoexcited charges (for $\Delta =0$ meV
and $\tau _{p}=2000$ ps, an excitation energy density of $1$ nJ cm$^{-2}$
corresponds to a characteristic density of \ $10^{10}$ cm$^{-2}$ in our
structure). It is important to point out that for such a long pulse, a
detuning energy of $\Delta =0.01$ meV is enough to reach the maximum of the
generation function $g(t)$ (Fig. 2)  \cite{26}. Thus, we will use this value
in calculations.


Since we will consider three photoexcitation cases, we will take into
account three possibilities: $\hbar \omega _{i}=(\varepsilon
_{c}-\varepsilon _{b,i}-\varepsilon _{v})+\Delta $ \ corresponding to the
laser energies $\hbar \omega _{i}$ ($i=e,exc,tr$) for free, excitonic and
trionic electrons, which resonate with their respective levels $\varepsilon
_{c}$, $\varepsilon _{c}-\varepsilon _{b,exc}$ and $\varepsilon
_{c}-\varepsilon _{b,tr}$, where $\varepsilon _{b,i}$ is the exciton or
trion binding energy (obviously $\varepsilon _{b,e}=0$). Thus, we will have
three generation functions $g_{i}(t)$ depending on the level resonance.
\bigskip

In the previous presentation of the density matrix we have not consider the
spin of the particles. Let us do. To study the time evolution of the
electron spin polarization, we first perform expressions for the temporal
evolution of free electron, hole, exciton, and negative trion densities,
paying attention to their spin orientation. We will consider the three
possible generation functions as well as all other processes of generation
and annihilation of these species.\ We will also consider the spin flip of
electrons and holes. Then, we will particularize these expressions to the
three cases under study.\bigskip

Initially, free electrons are injected into the QW, through the left
barrier, quasi-instantaneously and without any spin polarization. The
probability of spin $+1/2$ $\left( e\uparrow \right) $ or $-1/2$ $\left(
e\downarrow \right) $ is the same. Applying a circularly polarized light $%
\sigma ^{+}$ with energy $\hbar \omega _{e}=(\varepsilon _{c}-\varepsilon
_{v})+\Delta $, electron-hole pairs are generated with the peculiarity that
the hole angular momentum is $+3/2$ $\left( h\Uparrow \right) $ and that of
the electron is $-1/2$ $\left( e\downarrow \right) $  \cite{11}. Since the
spin flip time of the holes is very small in comparison with the duration
times of the remaining involved events, holes with angular momentum $-3/2$ $%
\left( h\Downarrow \right) $ almost immediately appear. As the time of spin
flip of electrons is several orders of magnitude greatest than the rest of
processes, it will only affect to the behavior of the system at very long
times. Nevertheless we will include it in our calculations.

The Coulomb interaction between electrons (injected or photoexcited) and
holes creates the exciton because the binding energy reduces the energy of
the system, leading to a more stable state. Initially all the holes, which
have angular momentum $+3/2\left( h\Uparrow \right) $, will join to
electrons $-1/2\left( e\downarrow \right) $ to give rise to excitonic bosons
with angular momentum $+1$ $\left( exc\Uparrow \downarrow \right) $. Almost
simultaneously, given the fast spin flip of the photoexcited holes, holes $%
-3/2$ $\left( h\Downarrow \right) $ and electrons $+1/2$ $\left( e\uparrow
\right) $ give rise to excitons with angular momentum $-1$ $\left(
exc\Downarrow \uparrow \right) $. We consider that nonoptical active
excitons with the same orientation of electron and hole spins cannot be
photoexcited or recombined. Inclusion of these cases in the Bloch equations
would lead to nonsense results.

The interaction of excitons with the remaining low-density electron gas
leads to the formation of trions. Excitons $\left( exc\Uparrow \downarrow
\right) ,$ in which the electron has spin $-1/2\left( e\downarrow \right) $,
together with free electrons of spin $+1/2\left( e\uparrow \right) $, form
trions (and conversely for the other orientation). The angular momentum of
these trions $\left( tr\Uparrow \text{, }tr\Downarrow \right) $ coincides
with the corresponding angular momentum of holes, because the electron spins
are compensated each other. Another possibility is the direct formation of
trions by the Coulomb interaction between electrons and holes without
intermediary pseudoparticles. The choices for angular momentum of these
trions are the same already mentioned.

When the circularly polarized light $\sigma ^{+}$ has energy $\hbar \omega
_{exc}=(\varepsilon _{c}-\varepsilon _{b,exc}-\varepsilon _{v})+\Delta $,
excitons are directly generated, with the same angular momentum presented
before, $+1$ $\left( exc\Uparrow \downarrow \right) $ with hole angular
momentum of $+3/2$ $\left( h\Uparrow \right) $ and electron angular momentum
of $-1/2$ $\left( e\downarrow \right) $. As in the former example, these
excitons can produce trions $+3/2$ $\left( tr\Uparrow \right) $ through
their interaction with free electrons $+1/2$ $\left( e\uparrow \right) $.

And, in the last case, when the energy of the polarized light resonates with
the trion level, $\hbar \omega _{tr}=(\varepsilon _{c}-\varepsilon
_{b,tr}-\varepsilon _{v})+\Delta $, the $+3/2$ $\left( tr\Uparrow \right) $
trions are generated by the capture of free electrons $+1/2$ $\left(
e\uparrow \right) $.

The above mentioned mechanisms take some time expressed through the
generation rates $F_{exc}$, $F_{tr}$, and $F_{tr2\text{ }}$. The inverse of
the mean time needed for exciton formation is $F_{exc}$, and $F_{tr}$ is the
trion formation rate via three-particle processes: two free electrons and a
hole for negative trion. And $F_{tr2\text{ \ }}$is the trion formation rate
through two-particle interaction: an exciton plus a free electron.

About the disappearance of the mentioned species, we will consider three
mechanisms. First, the direct recombination of a free electron with a hole.
Second, the recombination of an electron and a hole within an exciton, as
well as the dissociation of the exciton resulting in an electron and a hole.
And third, the recombination of an electron with a hole within a trion
leaving free the other electron, together with the dissociation of a trion
in two or three particles. In the following expressions, $R_{e}$ is the
radiative recombination coefficient of free carriers, $R_{exc}$ and $R_{tr}$
are the inverse of the intrinsic exciton and trion lifetimes, respectively.
Dissociation rates for both excitons and trions are so small that neglecting
their effect do not change results (Appendix A).

Moreover, electrons lose their free condition when producing excitons and
trions. So the generation rates $F_{exc}$, $F_{tr}$, and $F_{tr2\text{ }}$%
contribute also as three additional mechanisms for free electron extinction.
Further, trion annihilation gives rise to free electrons because the
recombination of one electron with the hole leaves an extra free electron.
Thus, we have another generation term for free electrons at a rate $R_{tr}$.
We also define $S_{e}$ and $S_{h}$ as the spin flip rate for electrons and
holes, respectively.

By writing densities of free electrons, holes, excitons, and negative trions
in units of $N_{ph}$ [$n_{e}(t),$ $n_{h}(t),$ $n_{exc}(t),$ and $n_{tr}(t)$,
respectively] we project over the two spin states. We will consider free
electrons from doping included in the initial boundary conditions, $%
n_{e}(t=0)=N_{D}/N_{ph}$ which, projected onto the spin states reads $\
n_{e\uparrow }(t=0)=n_{e\downarrow }(t=0)=N_{D}/2N_{ph}$, where $N_{D}$ is
the total density of injected electrons. The remaining densities are equal
to zero before switching the pulse on.

For free electrons we have 
\begin{eqnarray}
\frac{d}{dt}n_{e\uparrow }(t) &=&-F_{exc}n_{e\uparrow }(t)n_{h\Downarrow
}(t)-F_{tr2}n_{e\uparrow }(t)n_{ex\Downarrow \uparrow
}(t)-F_{tr}n_{e\uparrow }(t)n_{e\downarrow }(t)\left[ n_{h\Uparrow
}(t)+n_{h\Downarrow }(t)\right]  \notag \\
&&+R_{tr}n_{tr\Uparrow }(t)-R_{e}n_{e\uparrow }(t)n_{h\Downarrow }(t)+S_{e} 
\left[ n_{e\downarrow }(t)-n_{e\uparrow }(t)\right] -g_{tr}(t)n_{e\uparrow
}(t),  \notag \\
\frac{d}{dt}n_{e\downarrow }(t) &=&g_{e}(t)-F_{exc}n_{e\downarrow
}(t)n_{h\Uparrow }(t)-F_{tr2}n_{e\downarrow }(t)n_{ex\Uparrow \downarrow
}(t)-F_{tr}n_{e\uparrow }(t)n_{e\downarrow }(t)\left[ n_{h\Uparrow
}(t)+n_{h\Downarrow }(t)\right]  \notag \\
&&+R_{tr}n_{tr\Downarrow }(t)-R_{e}n_{e\downarrow }(t)n_{h\Uparrow
}(t)+S_{e} \left[ n_{e\uparrow }(t)-n_{e\downarrow }(t)\right] .  \label{10}
\end{eqnarray}%
In a similar way, excitonic electrons appear at a rate $F_{exc}$ and
disappear due to recombination and trion generation at rates $R_{exc}$ and $%
F_{tr2}$, respectively. Thus, for excitonic electrons, 
\begin{eqnarray}
\frac{d}{dt}n_{exc\Downarrow \uparrow }(t) &=&g_{exc}(t)+F_{exc}n_{h\Uparrow
}(t)n_{e\downarrow }(t)-R_{exc}n_{exc\Downarrow \uparrow
}(t)-F_{tr2}n_{exc\Downarrow \uparrow }(t)n_{e\uparrow }(t),  \notag \\
\frac{d}{dt}n_{exc\Uparrow \downarrow }(t) &=&F_{exc}n_{h\Downarrow
}(t)n_{e\uparrow }(t)-R_{exc}n_{exc\Uparrow \downarrow
}(t)-F_{tr2}n_{exc\Uparrow \downarrow }(t)n_{e\downarrow }(t)  \label{11}
\end{eqnarray}%
and, for up and down trion electrons,%
\begin{eqnarray}
\frac{d}{dt}n_{tr\Uparrow }(t) &=&g_{tr}(t)n_{e\uparrow
}(t)+F_{tr2}n_{e\uparrow }(t)n_{exc\Downarrow \uparrow
}(t)+F_{tr}n_{e\uparrow }(t)n_{e\downarrow }(t)n_{h\Uparrow
}(t)-R_{tr}n_{tr\Uparrow }(t)+S_{h}\left[ n_{tr\Downarrow }(t)-n_{tr\Uparrow
}(t)\right] ,  \notag \\
\frac{d}{dt}n_{tr\Downarrow }(t) &=&F_{tr2}n_{exc\Uparrow \downarrow
}(t)n_{h\downarrow }(t)+F_{tr}n_{e\downarrow }(t)n_{e\uparrow
}(t)n_{h\Downarrow }(t)-R_{tr}n_{tr\Downarrow }(t)+S_{h}\left[ n_{tr\Uparrow
}(t)-n_{tr\Downarrow }(t)\right] .  \label{12}
\end{eqnarray}%
Lastly, for the hole density we can write%
\begin{eqnarray}
\frac{d}{dt}n_{h\Uparrow }(t) &=&g_{e}(t)-F_{ex}n_{h\Uparrow
}(t)n_{e\downarrow }(t)-F_{tr}n_{e\uparrow }(t)n_{e\downarrow
}(t)n_{h\Uparrow }(t)-R_{e}n_{h\Uparrow }(t)n_{e\downarrow }(t)+S_{h}\left[
n_{h\Downarrow }(t)-n_{h\Uparrow }(t)\right] ,  \notag \\
\frac{d}{dt}n_{h\Downarrow }(t) &=&-F_{ex}n_{h\Downarrow }(t)n_{e\uparrow
}(t)-F_{tr}n_{e\uparrow }(t)n_{e\downarrow }(t)n_{h\Downarrow
}(t)-R_{e}n_{h\Downarrow }(t)n_{e\uparrow }(t)+S_{h}\left[ n_{h\Uparrow
}(t)-n_{h\Downarrow }(t)\right] .  \label{13}
\end{eqnarray}

To analyze particle densities we numerically perform the coupled system
(10-13) using the Runge-Kutta method. We only consider cases in which the
excess of free electron density is of about $10^{10}cm^{-2}$. For densities
higher than these, the effect of the electron-electron Coulomb interaction
becomes more remarkable because of the space-charge potential energy created
by the spatial distribution of electrons and holes. This space-charge
potential is repulsive for holes and attractive for electrons. At low
densities, electron-hole attraction dominates over electron-electron and
hole-hole repulsion. When the carrier density increases, the repulsive part
of the Hartree-Fock potential energy increases as well. Beyond a certain
initial density ($n_{e}(0)\gtrsim 10^{11}\ cm^{-2}$) of free electrons, the
repulsion equals the attractive potential. For higher densities the Coulomb
interaction of the second electron with the hole in the trion is canceled,
resulting in trion extinction, which creates excitons and free electrons. If
the free electron density increases further, the binding energy of excitons
tends to zero and these species also disappear. These densities mainly
affect the dynamics of the energy level shift of electrons \cite{3,26,27}.

To study the influence of the relation between doping and photoexcitation
energy density on the evolution of the different particle densities $%
n_{i}(t) $ (where $i=e\uparrow ,e\downarrow ,exc\Uparrow \downarrow
,exc\Downarrow \uparrow ,tr\Uparrow ,tr\Downarrow ,h\Uparrow ,$ $h\Downarrow 
$), we consider in calculations two photoexcitation characteristic densities
and two doping concentrations. To standardize equations, we take a referral
density $N_{0}$ as normalization constant for $n_{i}(t)$\ at any case. Then,
in a first case (A) we consider this referral density equal to the
photoexcited electron and the injected electron densities $\left(
N_{0}=N_{ph}=N_{D}\right) .$ This situation is reflected in the initial
boundary conditions stating $n_{e\uparrow }(t=0)=n_{e\downarrow }(t=0)=0.5$
in $N_{0}$ units. In a second case (B) we keep the same doping $\left(
N_{D}=N_{0}\right) $\ and use a $N_{ph}$ ten times bigger than the used
before $\left( N_{ph}=10N_{0}\right) $, which corresponds to an excitation
energy density ten times higher, approximately. The initial boundary
conditions do not change. In a last case (C) we return to the first
photoexcited electron density $\left( N_{ph}=N_{0}\right) $ but reduce the
injected electron density $\left( N_{D}=N_{0}/10\right) $.\ This means that
doping concentration is ten times lower in referral density units. So the
initial boundary conditions will be $n_{e\uparrow }(t=0)=$ $n_{e\downarrow
}(t=0)=0.05$.

There is a large and sparse number of values for the coefficients involved
in these equations (10-13), depending on the characteristics of the samples,
theories, or experimental conditions. These can be found in the related
literature, where there are many information available. Phenomenological and
theoretical data for electron spin relaxation and for GaAs-GaAlAs QWs are
included in\ many papers  \cite{29,30,31,32,33}. Also, data for hole spin
flip times can be found in the literature \cite{34,35,36}. An extensive
analysis of the formation, recombination and dissociation coefficients can
be found in the papers of Esser et al. \cite{1} and Portella-Oberli et al.%
 \cite{4,37}. Following these last references and considering the mass action
law (Saha-Eggert relations), we have evaluated the corresponding values for
our structure (Appendix A). Table I shows numerical values of the
coefficients used in this work. We have found the process of exciton
generation is much faster than the formation of trions and the annihilation
of both excitons and trions. Further, trion generation by means of excitons (%
$F_{tr2}$) is a mechanism faster than trion formation from three elements ($%
F_{tr}$), as expected. On the other hand, trion recombination is faster than
this last formation mechanism.

\begin{table}[tbp]
\begin{tabular}{|c|c|c|c|c|c|c|c|}
\hline
$R_{e}$ & $F_{exc}$ & $R_{exc}$ & $F_{tr2}$ & $F_{tr}$ & $R_{tr}$ & $S_{e}$
& $S_{h}$ \\ \hline
$10^{-4}$ & $2\times 10^{-2}$ & $2.5\times 10^{-3}$ & $6\times 10^{-3}$ & $%
2\times 10^{-3}$ & $4.5\times 10^{-3}$ & $10^{-4}$ & $2.6\times 10^{-2}$ \\ 
\hline
\end{tabular}%
\caption{Characteristic generation and relaxation coefficients in ps$^{-1}$,
for low temperature (5 K).}
\end{table}

Next, we calculate the electron spin relative polarization defined as the
difference between spin down and spin up free electron concentrations
divided by their sum, $\ p(t)=\left( n_{e\downarrow }(t)-n_{e\uparrow
}(t)\right) /\left( n_{e\downarrow }(t)+n_{e\uparrow }(t)\right) $.

Finally, we compare temporal evolution of normalized densities $n_{i}(t)$
and electron spin relative polarization $p(t)$, for the three above
mentioned cases of photoexcitation and doping (A, B and C), when resonant
photoexcited level corresponds to free, exciton or trion electron energy
level (cases 1, 2 and 3, respectively).

\section{Results and discussion}

\subsection{Case 1. Resonant photoexcitation of free electron.}

The spin polarization of the free electron gas takes place essentially by
the proper photoexcitation, because $\sigma ^{+}$ polarized light generates
spin down but not spin up electrons.

First, we will analyze the role that play the different processes on the net
polarization for the case (A). In the hypothetical case in which only doping
and photoexcitation coexist, without any other process, spin up electron
concentration would remain constant, whereas spin down electron
concentration linearly increases while the pulse is present (Fig. 3a), which
would produce certain polarization `per se'. These spin down and spin up
electron concentrations will be slightly affected by the inclusion of the
free electron-hole recombination along with the rapid flip of the hole spin
(Fig. 3b). They will change most appreciably for a realistic case, when all
the different processes of bimolecular creation and destruction of excitons
(Fig. 3c) or excitons and trions (Fig. 3d) are included.


The holes generated during the photoexcitation have spin up but, since the
hole flipping time is very short, it is possible to consider a similar
number of holes with spin up or down. However, due to the very slow flip of
the electron spin, the density of spin down electrons $\left( e\downarrow
\right) $ is greater because of the light polarization. As a consequence of
its higher density, free electron recombination is greater for these spin
down electrons. This could be explained by considering the situation where
spin up free holes $\left( h\Uparrow \right) $ can find easier this kind of
free electrons $\left( e\downarrow \right) $ to recombine with\ (Fig. 3b).

For the same reason, the disappearance of up and down electrons during the
exciton formation process is not balanced: spin down electrons form more
excitons $\left( exc\Uparrow \downarrow \right) $ than the others. Thus, a
clear decrease of spin down electron $\left( e\downarrow \right) $ density,
bigger than the corresponding for spin up electrons, can be observed by
comparing Fig. 3c with Fig. 3b. On the other hand, the recombination of
excitons is independent of the relative concentration for any spin
orientation because each electron in exciton is associated to a hole and its
recombination does not depend on others. Moreover, exciton recombination
does not influence both spin electron densities.

Since both electrons in trion have different spin orientation, three
particle trion formation leads to a balanced up and down electron quenching.
When the formation of trions arises from excitons, there are two
possibilities: excitons $\left( exc\Uparrow \downarrow \right) $ together
with spin up electrons $\left( e\uparrow \right) $ to give trions $\left(
tr\Uparrow \right) $ or excitons $\left( exc\Downarrow \uparrow \right) $
with spin down electrons $\left( e\downarrow \right) $ to give trions $%
\left( tr\Downarrow \right) $. As we mentioned above, there are a greater
amount of these electrons $\left( e\downarrow \right) $ and of the former
excitons $\left( exc\Uparrow \downarrow \right) $, leading to a balanced
formation of trions with both spin orientations. Thus, electrons with both
spin orientation disappear at a similar rate. Moreover, as the spin of
electrons in trion is compensated, the spin of the hole in trion is able to
flip and so, we can find trions with spin up and down with almost the same
probability. The same reasoning for exciton recombination applies to trions
recombination. In this case, the recombination of the electron-hole pair in
trion frees an electron, whose spin orientation is equiprobable due to the
similar density of the two kind of trions. In order to show all these
behaviors we have added Fig. 4, which corresponds to a magnified version of
Fig. 3d, with the curves split for the different spin orientations for
holes, excitons, and trions.


The resulting relative polarization $p(t)$ will reflect all the changes
above reported. In Fig. 5 we show the polarization calculated for each case
included in Fig. 3. Thus, in the case depicted in Fig. 3a, $n_{e\downarrow
}(t)$ increases linearly along the pulse duration giving the largest
difference of densities $\left( n_{e\downarrow }(t)-n_{e\uparrow }(t)\right) 
$. However, the relative polarization is not the best because also the total
amount of electrons $\left( n_{e\downarrow }(t)+n_{e\uparrow }(t)\right) $
is the biggest. The inclusion of the free electron- hole recombination is
not very noticeable in the concentrations behavior (Fig 3b) but it affects
to the polarization leading to the lowest values. Polarization increases
when the different processes of creation and annihilation of excitons are
included (corresponding to Fig. 3c),\ because the total density of electrons
with different spins is lower.\ And so, when also creation and destruction
of trions are considered (Fig 3d), the relative polarization reaches their
greatest values.


The influence of the photoexcitation energy density and the doping on the
different species normalized densities can be seen in Fig. 6, where we
compare the three cases mentioned before: (A) with $N_{ph}=N_{D}=$ $N_{0}$,
(B) with $N_{ph}=10N_{0}$, $N_{D}$ $=N_{0}$, and (C) with $N_{ph}=N_{0}$, $%
N_{D}$ $=N_{0}/10$. In the upper panel we represent case (A), which
corresponds to case considered in the former Figs. 3d and 4. The middle
panel displays case (B), where the same doping as in (A) but ten times
bigger characteristic photoexcitation density leads to a considerable
increase of excitons $\left( exc\Uparrow \downarrow \right) $, whose density
even exceeds that of the electrons $\left( e\downarrow \right) $. Now there
is also a significant trion concentration. The lower panel shows the last
case (C), where photoexcitation energy density goes back to the value of
(A), but doping is ten times lower. Although the relation between $N_{ph}$
and $N_{D}$ recovers the previous case ($N_{ph}=10N_{D}$), the behavior of
normalized densities $n_{i}(t)$ is clearly different. In this case, the
bigger influence of the photoexcitation leads to a big increase on the
relative density of spin down electrons $\left( e\downarrow \right) $ and
spin up holes $\left( h\Uparrow \right) .$ Now, the exciton $\left(
exc\Uparrow \downarrow \right) $ formation is not as faster as in (B) and a
greater variation in densities of spin up and spin down electrons is
obtained. As in (A), trion density is practically negligible.

\ 

Fig. 7 displays the resulting relative polarization for the three cases
considered. One can see the clear improvement of the electron spin
polarization when characteristic photoexcitation density exceeds doping
density ($N_{ph}=10N_{D}$, cases B and C). Moreover, relative polarization
is better when increasing $N_{ph}$ (B) than when diminishing $N_{D}$ (C).


\subsection{Case 2. Resonant photoexcitation of excitons.}

When excitons are photoexcited resonantly  \cite{23}, the only free electrons
that exist are those from the doped region (this is only strictly true at
low temperatures; if the temperature increases electrons and holes preferred
free carriers position). Moreover, because of all the holes are bound in
excitons, there are not free holes. Due to the circularly polarized light $%
\sigma ^{+}$, photoexcited excitons are $\left( exc\Uparrow \downarrow
\right) $ and their only alternative to form trions is to join spin up free
electrons $\left( e\uparrow \right) $\ from the doping, leading initially\
to specific trions $\left( tr\Uparrow \right) $. But, because of the fast
spin flip of the hole in trion we will find, after several picoseconds,
balanced densities for trions $\left( tr\Uparrow \right) $ and $\left(
tr\Downarrow \right) $. Thus, the recombination of an electron-hole pair in
these trions will leave spin up and spin down electrons with the same
probability. In this case, since there are not free holes to bind with free
electrons from doping, there is not the possibility of having excitons $%
\left( exc\Downarrow \uparrow \right) $; there will exist only photoexcited $%
\left( exc\Uparrow \downarrow \right) $ ones.

Fig. 8 shows the density time evolution for the different species and for
the same three cases of Fig. 6. As expected, resonance with free electron or
exciton level leads to a faster increase of the relative density of spin
down electrons $\left( e\downarrow \right) $ or excitons $\left( exc\Uparrow
\downarrow \right) $, respectively. The difference between the two kinds of
resonant photoexcitation is more obvious in lower panels (B and C), when $%
N_{ph}=10N_{D}$. By comparing upper panel (corresponding to case A) in both
figures, a smaller difference between spin up and down free electron
densities can be seen for this exciton resonance case. This is more evident
as we go through cases B and C. Accordingly, this kind of resonant
photoexcitation does not present any improvement for spin polarization. Fig.
9 depicts the relative electron spin polarization, showing a clear
improvement when increasing the photoexcitation energy density (case B), as
it occurred formerly (Fig. 7). However, for the other cases (A and C), spin
polarization does not increase as much as when light resonated with free
electrons.



\subsection{Case 3. Resonant photoexcitation of trions.}

Finally, we analyze the case of resonant photoexcitation with the trion
level. In this case, trions $\left( tr\Uparrow \right) $ are directly
generated by means of the capture of free electrons $\left( e\uparrow
\right) $ and without exciton intervention  \cite{38}. As in the previous
case, the unique free electrons we have are the injected ones, and there are
not free holes. So, as there are not free holes to join injected free
electrons, there is not the possibility of having excitons at all. We will
find just photoexcited trions and free electrons. The circularly polarized
light $\sigma ^{+}$ initially\ photoexcites specific trions $\left(
tr\Uparrow \right) $ and, due to the fast spin flip of the hole in trion, we
will have balanced densities for trions $\left( tr\Uparrow \right) $ and $%
\left( tr\Downarrow \right) $, as in the exciton resonant case.

Fig. 10 presents the evolution of the normalized density for the three\
situations under study. Due to the balanced densities of the two
orientations, trions $\left( tr\Uparrow \right) $ and $\left( tr\Downarrow
\right) $, we have drawn the sum of both curves in only one. While in the
first case (A) behavior is very similar to the former ones (Figs. 6 and 8),
does not happen the same when $N_{ph}=10N_{D}$ (B and C). In case (B), with $%
N_{ph}=10N_{0}$, $N_{D}$ $=N_{0}$, we can see a fast increase of the
relative density of trions, as expected for the photoexcited species but it
is clearly less than that shown in Figs. 6-8. The reason is that now, to
generate trions, there must be free electrons available to be captured.
Since the only electrons we have came from the injected from the doped
region, this electron density determines the possible photoexcited trion
density. Also, because only up free electrons $\left( e\uparrow \right) $
are captured during photoexcitation, the density of these electrons
drastically decreases. However, trion recombination leaves a compensated
density of \ up and down free electrons.

For $N_{ph}=N_{0}$, $N_{D}$ $=N_{0}/10$ (C), differences are more remarkable
among this and the two other cases (Figs. 6 and 8). Low panel in Fig. 10
looks like the upper one, except for the density axis. The key is that, now,
doping is very low and limits trion generation. Thus, in this case we do not
find a significative improvement on the density of photoexcited species.

Fig. 11 shows electron spin relative polarization for the three situations
under study, and when photoexciting in resonance with trion level. As
expected, curves coincide for cases A and C. By comparing this figure with
Figs. 7 and 9, we can note that, in the present case, spin polarization is
substantially worse than in the others, being free electron resonant
photoexcitation which leads to the best results.


\section{Summary and conclusions}

First, we should mention here the approximations used in the method of
calculation. We assume injection as a tunneling process much shorter than
all the other processes involved. Thus, we include doping in the initial
conditions as prior to photoexcitation. An important point is the possible
effect of the application of external electric fields on trion and electron
orientation. We used field strengths less than 10 kV/cm to prevent the
possible ionization or diffusion of trion through the structure   \cite{3}. \
Higher external fields can affect spin relaxation times through the
Elliott--Yafet mechanism  \cite{11}.

The fact that excitonic and trionic electron levels are very close in
energy, less than $2$ meV in the GaAs  \cite{9}, should be reflected in a
nonzero trion generation rate when excitons are photoexcited. Thus, a few
electrons should be generated in the corresponding state. In the present
study the amount of these electrons is negligible because the long pulse
leads to a narrow energy dispersion. In this way we have only considered
trions formed by means of excitons and free electrons because direct trion
generation by resonant exciton photoexcitation can be neglected.

We have used in numerical calculations the eight-coupled Bloch system
obtained from the kinetic equation for the one-electron density matrix,
projected over the spin states, and taking into account eight possible
time-dependent processes.\bigskip

In summary, we have analyzed the temporal behavior of the density of free
electrons, excitons and trions in doped QWs when they are photoexcited with $%
\sigma ^{+}$ circularly polarized light. We pay special attention to the
electron spin orientation when the sample has different doping
concentrations and is subjected to different photoexcitation energy
densities. We have considered three cases of resonant photoexcitation
corresponding to the three lower conduction electronic levels (free,
excitonic and trionic electrons). Also, we have studied the electron spin
relative polarization considering the role played by excitons and trions.
Our main conclusion is that, for practical purposes, photoexcitation of free
electron level leads to a higher spin polarization. Another important point
is the relationship between the characteristic photoexcitation density
(directly related to the photoexcitation energy density) and doping density.
The effect of both variables on the spin polarization is opposite, being
significantly more pronounced for the first one. Thus, we have found that
increasing photoexcitation enhances spin polarization while this improvement
can be achieved by reducing the doping. Our results show a wide variety of
responses, caused by the different carrier densities of free and bound
electrons and different photoexcitation energy densities. Present results
can be checked by means of photoluminescence measures. Photoluminescence
experiments of spin density in $n$-type GaAs-GaAlAs QWs are available  \cite%
{33}. This work can be applied to any spin relaxation mechanism and other
structures by changing the parameters involved. We expect that this work
will aid in the design of the experimental conditions to study electron spin
dynamics, as well as in stimulating research in this field.

\appendix

\section{Coefficients for different processes}

Let us check coefficients involved in the mechanisms with which we have been
working on: formation ($F_{i}$), dissociation ($D_{i}$), and recombination ($%
R_{i}$) of excitons and trions. Essentially, we have followed the approach
of Berney et al.  \cite{37}, specifying as far as possible to our structure
and $T=5$ K. Thus, we can outline processes as follows:%
\begin{eqnarray}
e^{-}+h^{+} &\rightleftarrows &X  \notag \\
X+e^{-} &\rightleftarrows &X^{-}  \notag \\
2e^{-}+h^{+} &\rightleftarrows &X^{-}.  
\end{eqnarray}%
We first consider the formation coefficients, which depend on carrier and
the lattice temperatures, $T_{c}$ and $T$, respectively. For the exciton
coefficient $C$ we have extrapolated a value of $\ 2\times 10^{-12}$ cm$^{2}$%
ps$^{-1}$ from Fig. 1 in the paper of Berney et al. Considering \ our
coefficient $F_{exc}=CN_{0}$, we obtain a rate of $F_{exc}\sim 2\times
10^{-2}$ ps$^{-1}$. In order to estimate bi- and tri-molecular trion
formation coefficients we have followed Fig. 2 of the same paper, obtaining $%
A_{2}\sim 0.6$ $\times 10^{-12}$ cm$^{2}$ps$^{-1}$and $A_{3}\sim 0.2\times
10^{-22}$ cm$^{4}$ps$^{-1}$. Thus, formation rate coefficients we have used
for trion formation are $F_{tr2}=A_{2}N_{0}\sim 6$ $\times 10^{-3}$ ps$^{-1}$
and $F_{tr3}=A_{3}N_{0}^{2}\sim 2$ $\times 10^{-3}$ ps$^{-1}$.

Coefficients $D_{i}$ have been calculated using the Saha-Eggert equations.
In the case that concerns us, Saha-Eggert relations can be written as%
\begin{eqnarray*}
\frac{n_{e}n_{h}}{n_{exc}} &=&K_{exc}(T)=\frac{m_{e}m_{h}}{m_{exc}}\frac{%
k_{B}T}{2\pi \hbar ^{2}}\exp (-\frac{E_{exc}}{k_{B}T}), \\
\frac{n_{e}n_{exc}}{n_{tr}} &=&K_{tr2}(T)=\frac{m_{e}m_{exc}}{m_{tr}}\frac{%
k_{B}T}{2\pi \hbar ^{2}}\exp (-\frac{E_{tr}-E_{exc}}{k_{B}T}), \\
\frac{n_{e}^{2}n_{h}}{n_{tr}} &=&K_{tr3}(T)=\frac{m_{e}^{2}m_{h}}{m_{tr}}%
\left( \frac{k_{B}T}{2\pi \hbar ^{2}}\right) ^{2}\exp (-\frac{E_{tr}}{k_{B}T}%
),
\end{eqnarray*}%
where $E_{exc}=8.60$ meV and $E_{tr}=1.33$ meV are the binding energy for
exciton and trion, respectively  \cite{9}. Here $m_{exc}=m_{e}+m_{h}$, $%
m_{tr}=2m_{e}+m_{h},$where $m_{e}$, $m_{h}$ are the electron and hole
effective masses. Thus, we have obtained $K_{exc}=8.04$ cm$^{-2}$, $%
K_{tr2}=2.04\times 10^{2}$cm$^{-2}$, and $K_{tr3}=7.94\times 10^{17}$cm$%
^{-4} $. Dissociation coefficients will be $D_{exc}=CK_{exc}$, $%
D_{tr2}=A_{2}K_{tr2}$, and $D_{tr3}=A_{3}K_{tr3}$, which numerical values
are $D_{exc}=1.6\times 10^{-11}$ ps$^{-1}$, $D_{tr2}=1.2\times 10^{-10}$ ps$%
^{-1}$, and $D_{tr3}=1.6\times 10^{-5}$ ps$^{-1}$.

As we can see, the dissociation rates obtained are negligible compared with
the formation and recombination rates. Therefore, we have neglected the
dissociation coefficients in the system of equations (10-13) because they do
not bring any noticeable change in the results. \ \ \ \ \ \ \ 

Let us consider the radiative decay times for excitons and trions, which are
the inverse of the recombination coefficients $R_{i}$. As the above
coefficients, recombination rates also depend on carrier temperature  \cite{4}%
. Values for our structure and conditions are $R_{exc}=\tau _{exc}^{-1}=2.5$ 
$\times 10^{-3}$ ps$^{-1}$ and $R_{tr}=\tau _{tr}^{-1}=4.5$ $\times 10^{-3}$
ps$^{-1}$ for exciton and trion, respectively.

Now, we will consider \ free electron-hole radiative recombination. Unlike
exciton or trion recombination rates, which do not directly depend on their
concentrations, free electron recombination does. Free electron must find a
free hole to recombine while these charges are already bound in exciton and
trion. Following Szczytko et al.  \cite{39}, the bimolecular plasma
recombination rate will be $B\sim 10^{-14}$ cm$^{2}$ps$^{-1}$ and the
radiative recombination coefficient of free carriers, $R_{e}=BN_{0}\sim
10^{-4}$ ps$^{-1}$.

Finally, electron spin relaxation time was taken from Dzhioev et al.  \cite{33}%
, which empirically obtained a value of $\tau _{s}=10$ ns for a structure
and conditions similar to ours. About holes, we have used the spin flip time
obtained by Schneider et al.  \cite{34}, where $\tau _{h}=39$ ps. Thus, the
spin relaxation coefficients included in calculations are $S_{e}=\tau
_{s}^{-1}=10^{-4}$ ps$^{-1}$ for electrons and $S_{h}=\tau _{h}^{-1}=2.6$ $%
\times 10^{-2}$ ps$^{-1}$ for holes.

\newpage 
\begin{figure}[tbp]
\begin{center}
\includegraphics{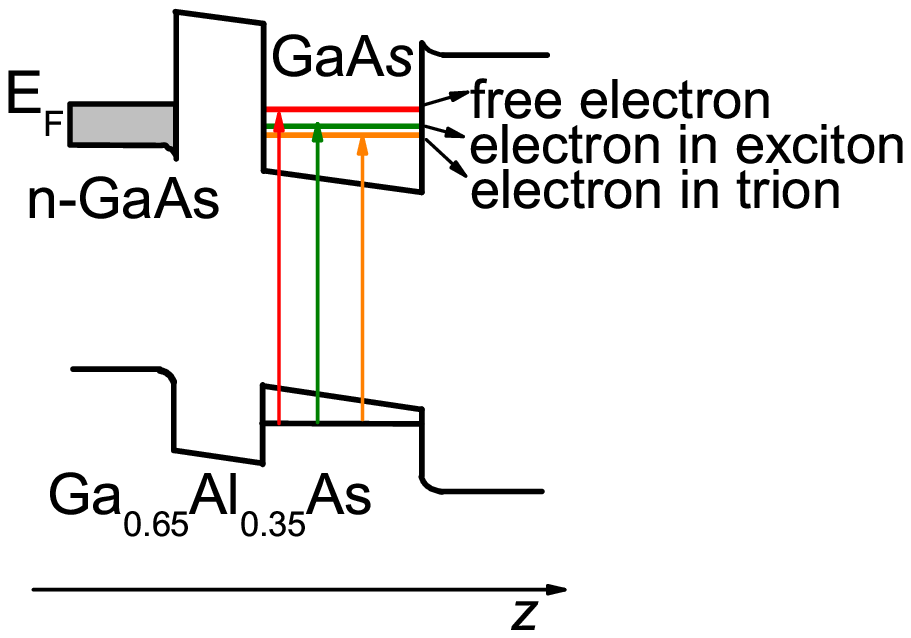}
\end{center}
\par
\addvspace{-2 cm}
\caption{(Color online) Scheme of the doped quantum well under study.}
\end{figure}

\newpage 
\begin{figure}[tbp]
\begin{center}
\includegraphics{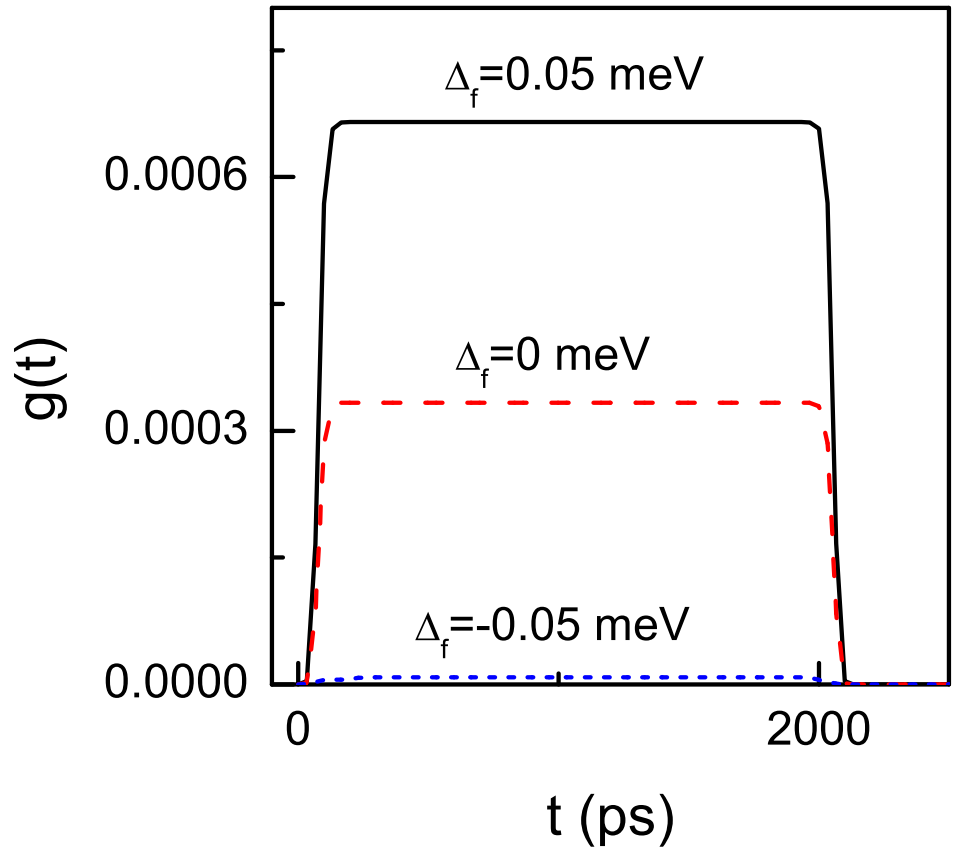}
\end{center}
\par
\addvspace{-2 cm}
\caption{{}(Color online) Generation function for $\protect\tau _{p}=2000$
ps and $\protect\tau _{f}=10$ ps. Solid line: $\Delta =0.05$ meV. Dashed: $%
\Delta =0$ meV. Dotted: $\Delta =-0.05$ meV. }
\end{figure}

\newpage 
\begin{figure}[tbp]
\begin{center}
\includegraphics{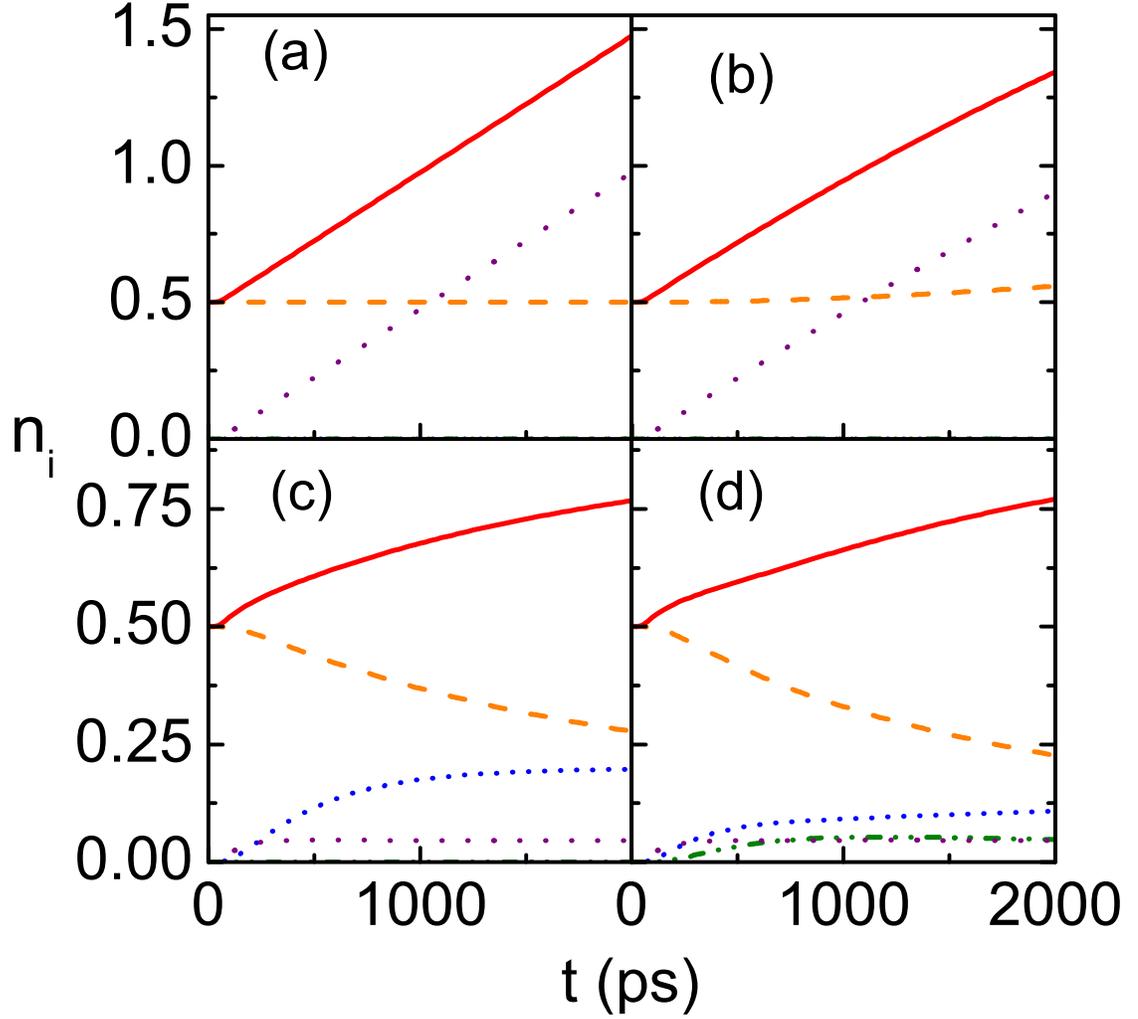}
\end{center}
\par
\addvspace{-6 cm}
\caption{(Color online) Normalized density versus time for free electron
resonant photoexcitation. Considering only doping and photoexcitation (a),
adding recombination electron-hole and spin-flip (b), considering also
formation and recombination of excitons (c), and including the presence of
trions (d). Solid line (red): spin-down free electrons; dashed line
(orange): spin-up free electrons; dash-dot-dotted line (purple): up and down
free holes; short-dashed line (blue): excitons; and dash-dotted line
(green): trions.}
\end{figure}

\newpage 
\begin{figure}[tbp]
\begin{center}
\includegraphics{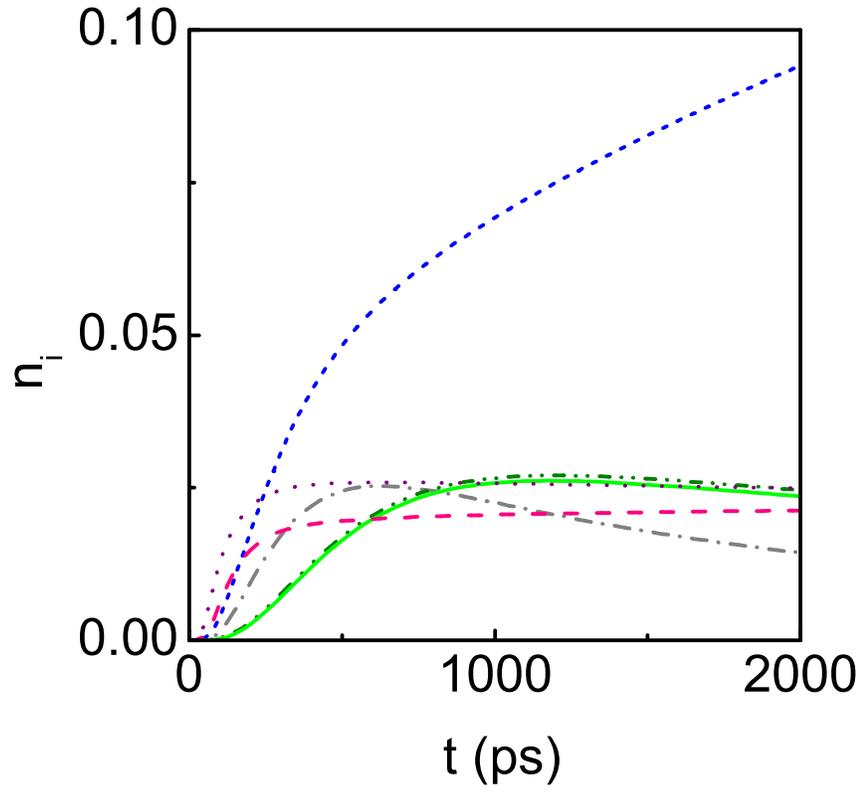}
\end{center}
\par
\addvspace{-2 cm}
\caption{(Color online) Normalized density versus time. The same as in Fig
3d but amplifying the region for excitons, trions and holes. Short-dashed
line (blue): excitons $\Uparrow \downarrow $; dash-dotted line (grey):
excitons $\Downarrow \uparrow $; dash-dot-dotted line (green): trions $%
\Uparrow $; solid line (light green): trions $\Downarrow $; dotted line
(purple): holes $\Uparrow $; and dashed line (pink) holes $\Downarrow $. }
\end{figure}

\newpage 
\begin{figure}[tbp]
\begin{center}
\includegraphics{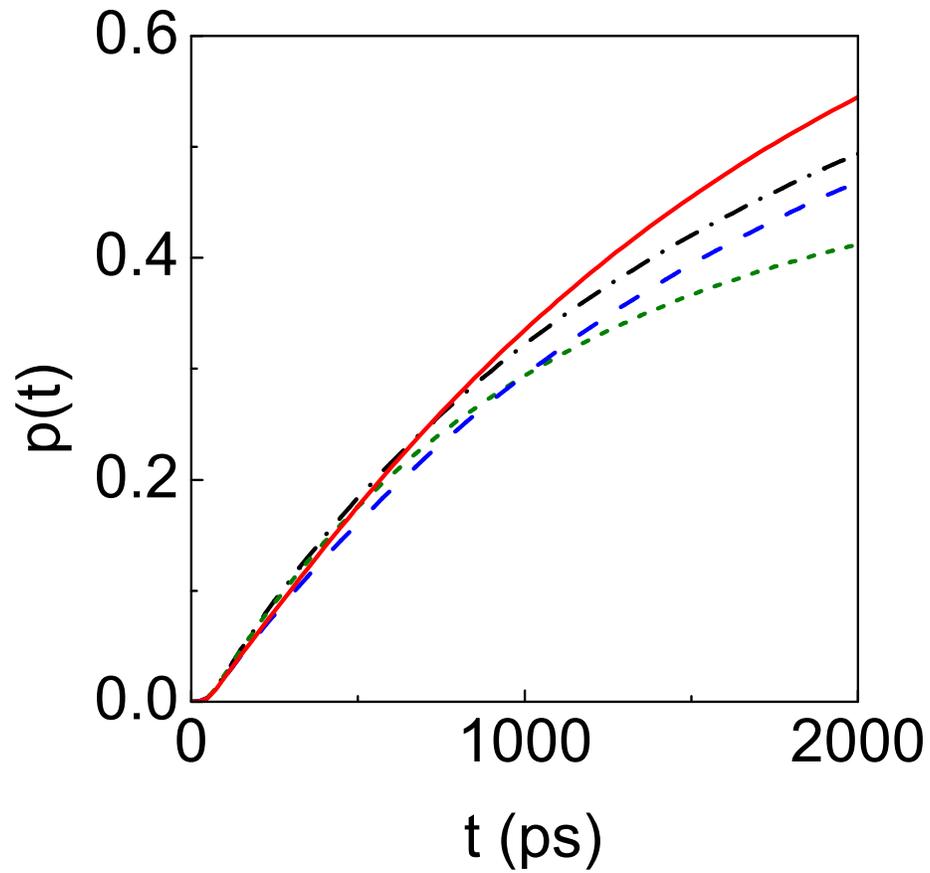}
\end{center}
\par
\addvspace{-2 cm}
\caption{(Color online) Polarization versus time. Considering only doping
and photoexcitation (dash-dotted black line), adding recombination
electron-hole and spin-flip (dotted green line), considering also formation
and recombination of excitons (dashed blue line), and including the
formation of trions (solid red line). }
\end{figure}

\newpage 
\begin{figure}[tbp]
\begin{center}
\includegraphics{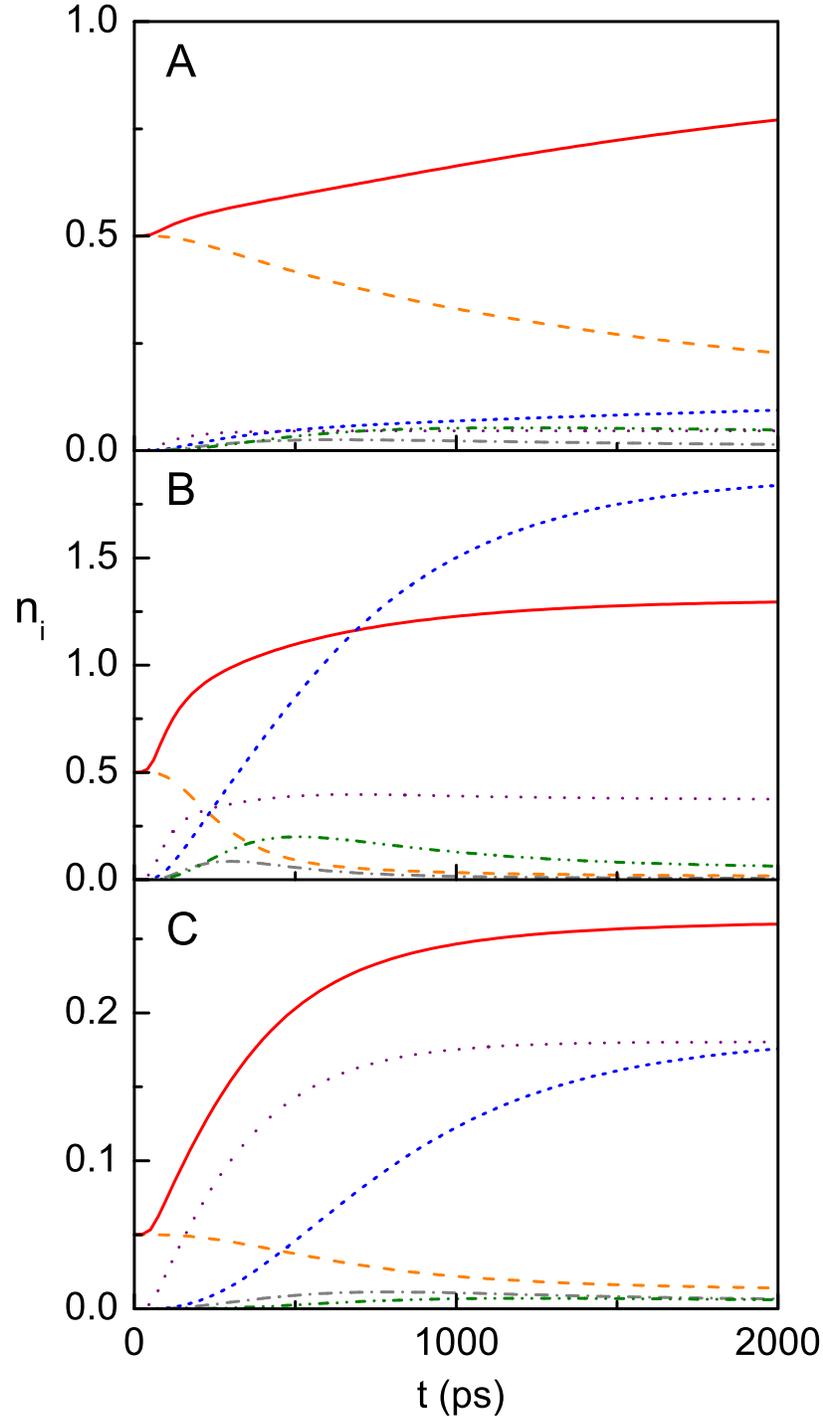}
\end{center}
\par
\addvspace{-2 cm}
\caption{(Color online) Normalized density time evolution for free electron
resonant photoexcitation. Upper panel: case (A), $N_{ph}=N_{D}$ $=N_{0}$;
middle panel: case (B), $N_{ph}=10N_{0}$, $N_{D}=N_{0}$; and \ lower panel:
case (C), $N_{ph}=N_{0}$ $N_{D}=N_{0}/10$. Colors and lines as in Figs. 3
and 4. }
\end{figure}

\newpage 
\begin{figure}[tbp]
\begin{center}
\includegraphics{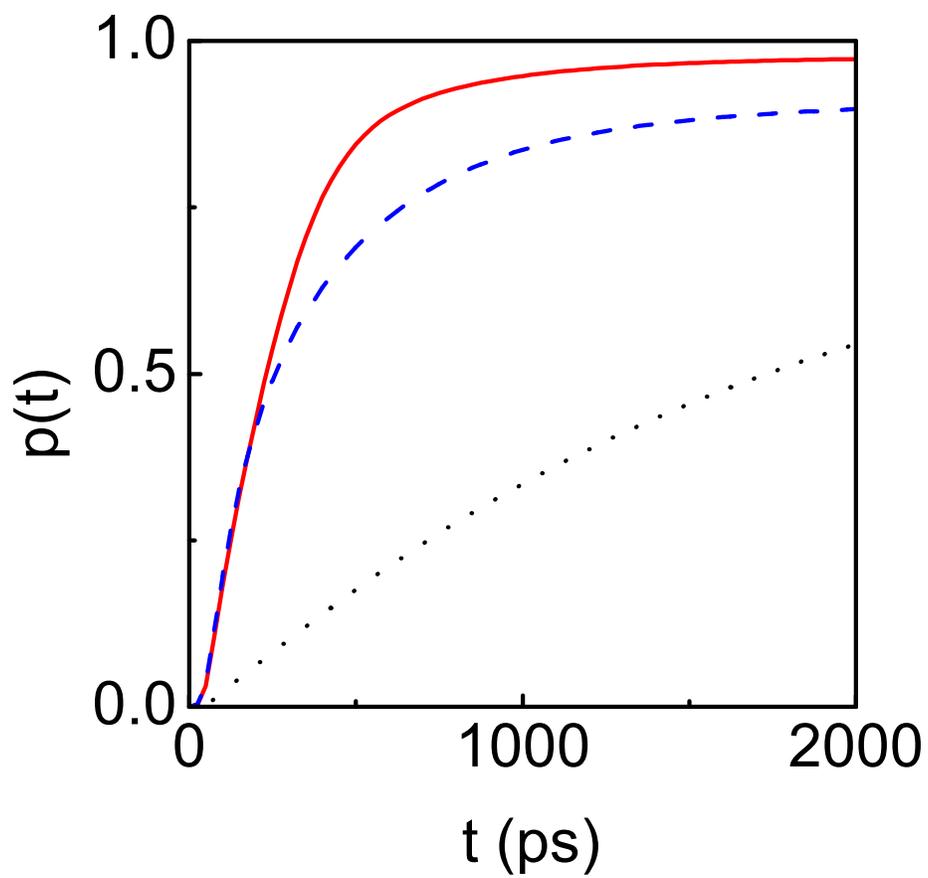}
\end{center}
\par
\addvspace{-2 cm}
\caption{(Color online) Relative polarization versus time. Case (A): dotted
(black) line: case (B): solid (red) line;\ and case (C): dashed line (blue).}
\end{figure}

\newpage 
\begin{figure}[tbp]
\begin{center}
\includegraphics{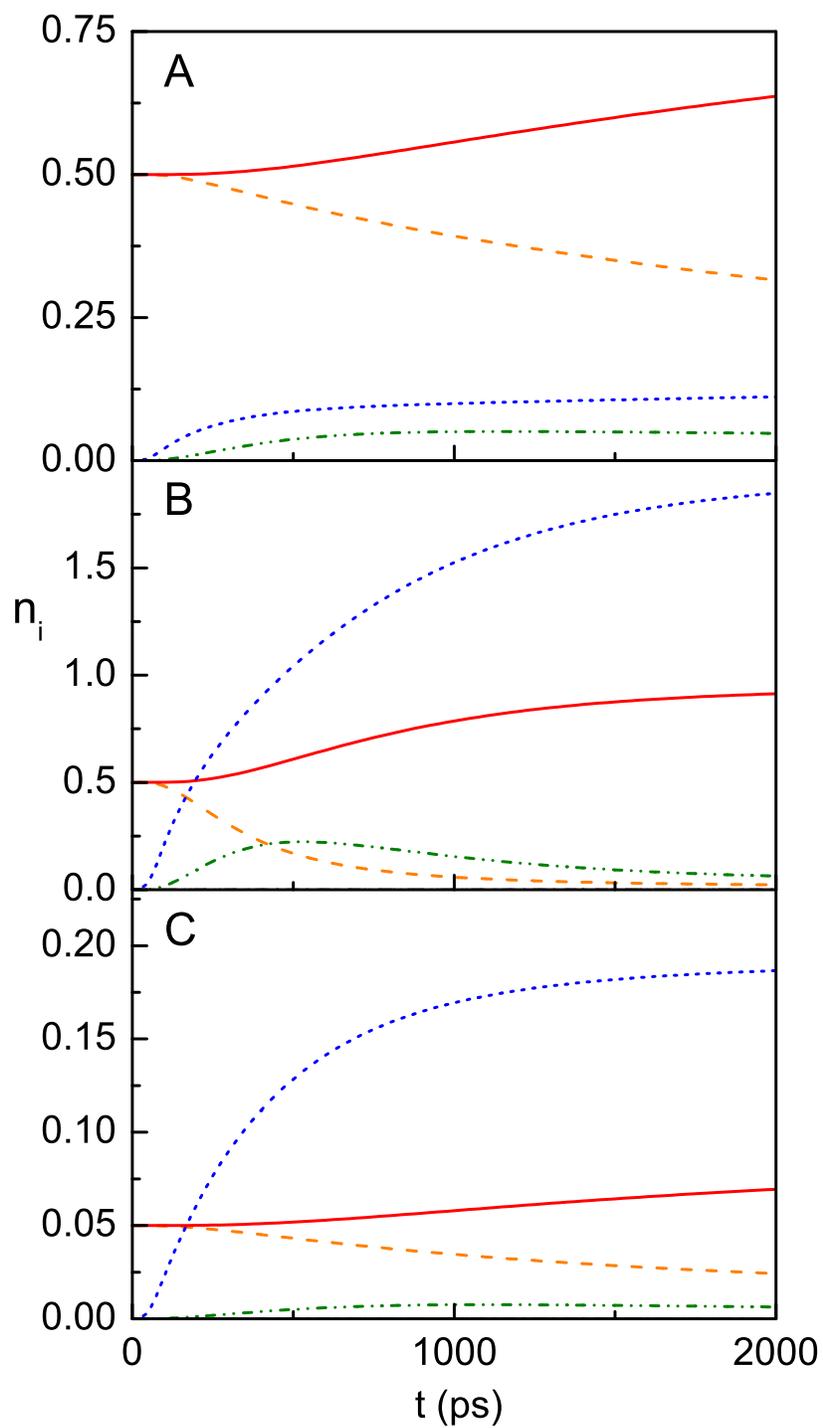}
\end{center}
\par
\addvspace{-2 cm}
\caption{(Color online) Normalized density versus time for exciton level
resonant photoexcitation. Upper, middle and lower panels correspond to the
same cases as in Fig. 6, with the same meaning for colors and lines. }
\end{figure}

\newpage 
\begin{figure}[tbp]
\begin{center}
\includegraphics{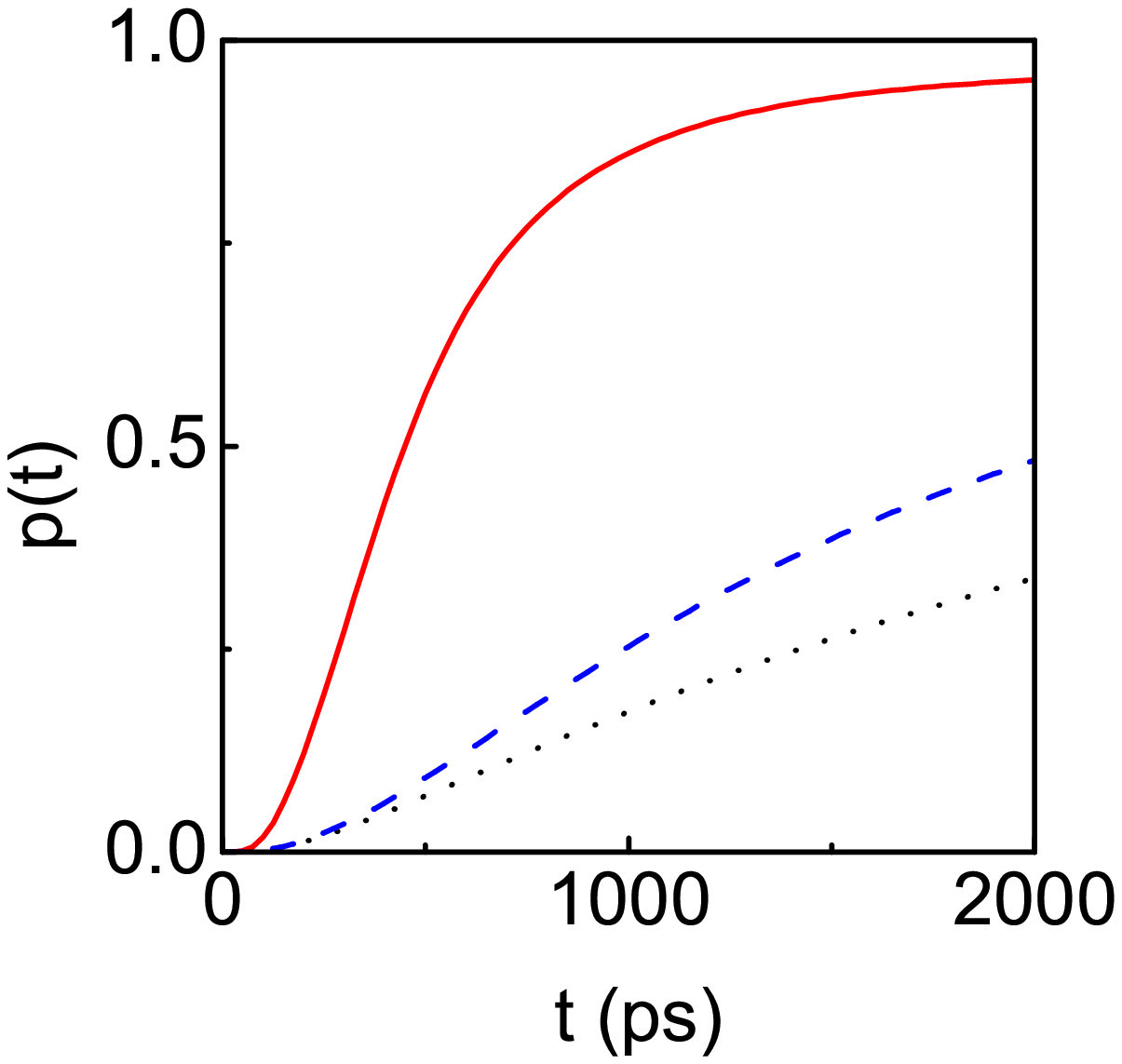}
\end{center}
\par
\addvspace{-2 cm}
\caption{(Color online) Electron spin polarization versus time for exciton
resonant\ photoexcitation. Colors and lines as in Fig 7. }
\end{figure}

\newpage 
\begin{figure}[tbp]
\begin{center}
\includegraphics{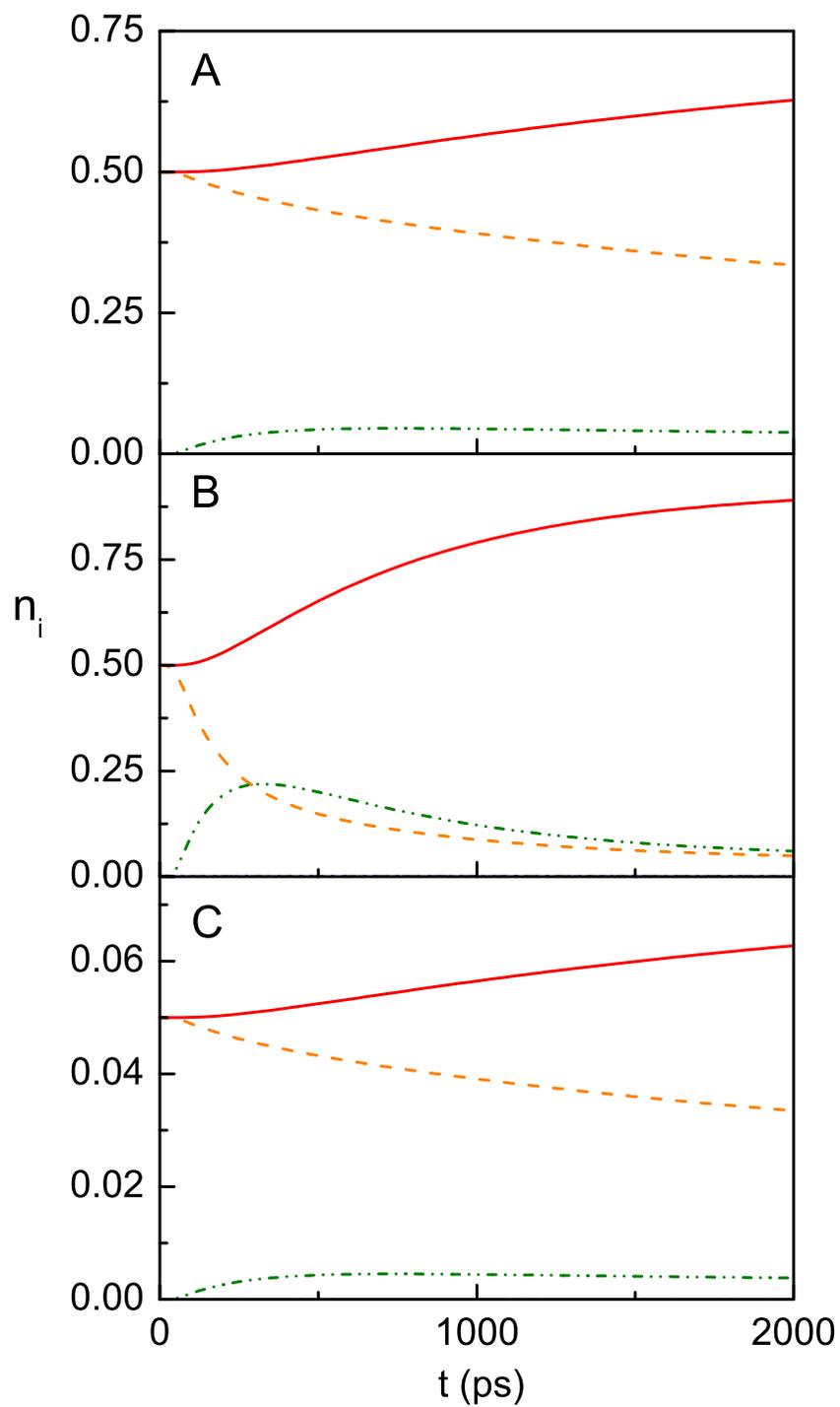}
\end{center}
\par
\addvspace{-2 cm}
\caption{(Color online) Relative density versus time for trion level
resonant photoexcitation. Upper, middle and lower panels for the same cases
as Fig. 6, with the same meaning for colors and lines.}
\end{figure}

\newpage 
\begin{figure}[tbp]
\begin{center}
\includegraphics{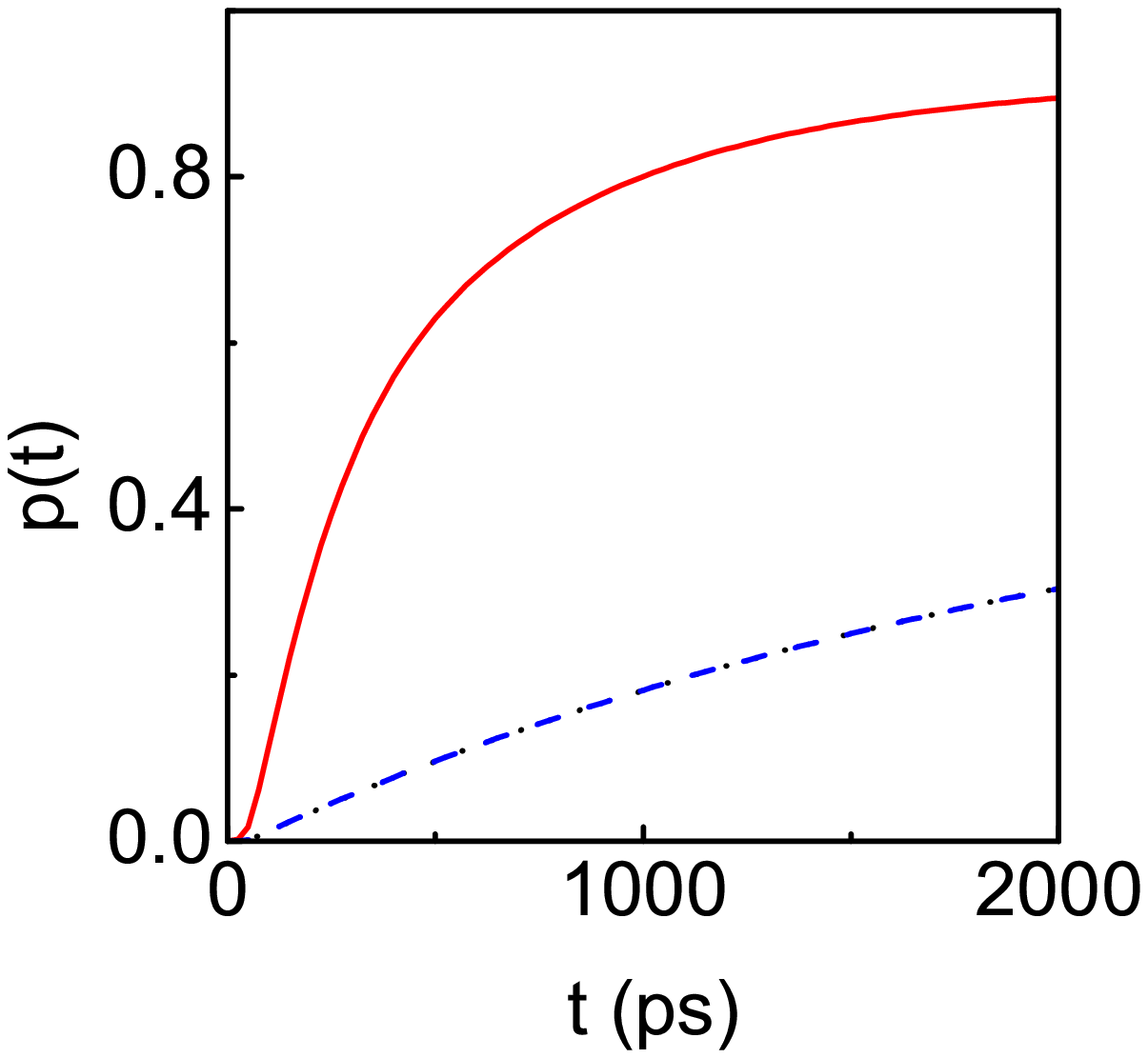}
\end{center}
\par
\addvspace{-2 cm}
\caption{(Color online) Electron spin polarization versus time for trion
resonant photoexcitation. Colors and lines as in Fig 7. }
\end{figure}

\end{document}